\documentclass{article}
\usepackage{waspaa21,amsmath,graphicx,times}
\usepackage{color}
\usepackage[binary-units=true]{siunitx}
\usepackage{comment}
\usepackage{booktabs}
\usepackage{listings}
\usepackage[hidelinks]{hyperref}

\newcommand{\kbps}[1]{$\SI[per-mode=symbol,per-symbol = /]{#1}{\kilo\bit\per\second}$}
\newcommand{\flops}[1]{$\SI[per-mode=symbol,per-symbol = /]{#1}{GMACs}$}

\title{A Streamwise GAN Vocoder for Wideband Speech Coding at Very Low Bit Rate}

\name{Ahmed Mustafa,
      Jan Büthe\sthanks{This work was done while the author was at Fraunhofer. Now he works at Fantasmo, \textit{email: Jan.Buethe@gmx.net}},
      Srikanth Korse,
      Kishan Gupta,
      Guillaume Fuchs,
      Nicola Pia}
\address{
Fraunhofer IIS, Erlangen, Germany\\
\{ahmed.mustafa.ahmed, srikanth.korse, kishan.gupta, guillaume.fuchs, nicola.pia\}@iis.fraunhofer.de
}

\begin{document}

\ninept
\maketitle

\begin{sloppy}

\begin{abstract}
  Recently, GAN vocoders have seen rapid progress in speech synthesis, starting to outperform autoregressive models in perceptual quality with much higher generation speed. 
However, autoregressive vocoders are still the common choice for neural generation of speech signals coded at very low bit rates. 
In this paper, we present a GAN vocoder which is able to generate wideband speech waveforms from parameters coded at \kbps{1.6}.  
The proposed model is a modified version of the StyleMelGAN vocoder that can run in frame-by-frame manner, making it suitable for streaming applications. 
The experimental results show that the proposed model significantly outperforms prior autoregressive vocoders like LPCNet for very low bit rate speech coding, with computational complexity of about \flops{5}, providing a new state of the art in this domain.
Moreover, this streamwise adversarial vocoder delivers quality competitive to advanced speech codecs such as EVS at \kbps{5.9} on clean speech, which motivates further usage of feed-forward fully-convolutional models for low bit rate speech coding.

\end{abstract}

\begin{keywords}
GAN vocoder, StyleMelGAN, neural speech synthesis, LPCNet, speech coding
\end{keywords}

\section{Introduction}
\label{sec:intro}

Despite decades of extensive work classical speech coders offer very low quality at bit rates under \kbps{3}.
New techniques based on the use of neural networks showed breakthrough advancements in this area in recent years, enabling compression factors much higher than conventional approaches, while maintaining acceptable quality.
Neural speech coders are based on the classical encoder-decoder scheme: the encoder analyzes the input signal and extracts a set of acoustic features, which are then quantized, coded and transmitted; the decoder reconstructs the input signal using the information contained in the received bit stream.
In neural speech coders a generative neural network plays the role of the decoder (i.e., neural vocoder), as illustrated in Figure~\ref{fig:neural_spco}.
It was demonstrated~\cite{wavenet_coding,valin2019lpcnetspco} that conditioning a neural vocoder with coded acoustic parameters could produce natural wideband speech at bit rates lower than \kbps{2}.

In recent years neural vocoders~\cite{wavenet,waveglow,pmlr-v80-kalchbrenner18a,lpcnet} have revolutionized fields such as text-to-speech, voice conversion and speech enhancement, generating speech of unprecedented high quality.
Most of these solutions however, are not suitable for speech coding purposes.
This is mainly due to their high computational complexity or very slow generation speed, with clear quality degradation when using coarsely quantized conditioning features.

\begin{figure}[t]
  \centering
  \centerline{\includegraphics[width=\columnwidth]{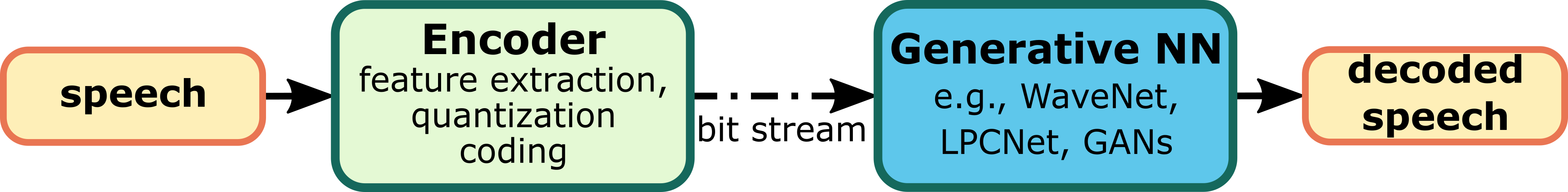}}
  \caption{High-level block-diagram of a neural speech coder.}
  \label{fig:neural_spco}
\end{figure}

Neural vocoders based on generative adversarial networks (GANs)~\cite{gans} were recently shown to be competitive and viable alternatives to autoregressive and flow-based models for speech synthesis applications~\cite{melgan,NEURIPS2020_c5d73680,smgan}.
However, they are by design not suited for streaming or real-time speech communication, since they take the advantage of heavy parallelization for processing large blocks of conditioning information at once.
This permits efficient generation of speech waveforms in one shot, but exploits the advantage of having the acoustic features encoding information about future samples, which are not available in a streaming scenario because of the high algorithmic delay they would cause. 
Moreover, GAN vocoders work particularly well with homogeneous speech representations such as mel-spectrograms, whereas speech coding applications primarily use non-homogeneous (e.g., parametric) speech representations that may not easily condition GAN vocoders for high-quality signal generation. 

To solve the above-mentioned issues, our contributions in this work are twofold:
\begin{itemize}
\item We propose Streamwise StyleMelGAN (SSMGAN), a modified StyleMelGAN vocoder for frame-by-frame generation of wideband speech at low delay, with reasonable computational complexity.
\item We demonstrate that SSMGAN is able to generate high-quality speech even when conditioned with a parametric and highly compressed representation provided by the encoder of LPCNet~\cite{valin2019lpcnetspco}, which delivers a \kbps{1.6} bitstream to our StyleMelGAN-based vocoder.
\end{itemize}

\section{Related Works}
\label{sec:rel_works}

The research on neural vocoders is a very active field with new models being presented every few months.
For this reason, here we only refer to some of the ones which sparked the most attention.
The first family to appear was the one of autoregressive models~\cite{wavenet,pmlr-v80-kalchbrenner18a,lpcnet}, followed by flow-based models~\cite{waveglow}, and then GANs~\cite{melgan,pwgan,Binkowski2020High,NEURIPS2020_c5d73680,smgan}.

The first work to show the feasibility of low bit rate neural speech coding was~\cite{wavenet_coding}, using a WaveNet decoder.
The decoder network's complexity makes it impossible to deploy it in concrete applications.
The complexity issue was partially tackled with a different approach in~\cite{samplernn_coding}.
Finally the LPCNet model~\cite{valin2019lpcnetspco} introduced optimizations which made neural speech coding possible on edge device.
Moreover, the coding scheme used in LPCNet has a very low bit rate of \kbps{1.6}.
The coding parameters include acoustic features classically used in parametric speech coding, i.e. the Bark scale cepstrum, the pitch information and the energy.
Table~\ref{tab:lpcnet_qfeatures} describes in detail these parameters and the bit budget allocated to code them.

\begin{table}[h] 
  \centering
\begin{tabular}{ll}
  \toprule
       Coding Parameter & Bits/packet \\
  \midrule
       Pitch lag & 6 \\
       Pitch modulation & 3 \\
       Pitch correlation & 2 \\
       Energy           & 7 \\
       Cepstrum absolute coding & 30\\
       Cepstrum delta coding & 13\\
       Cepstrum interpolation & 3\\
  \addlinespace
        Total & 64\\
  \bottomrule
\end{tabular}
\caption{LPCNet coding parameters and their bit allocation for a $\SI{40}{\milli\sec}$ packet}
\label{tab:lpcnet_qfeatures}
\end{table}

LPCNet's \kbps{1.6} decoder is an autoregressive architecture based on WaveRNN generating sample-by-sample wideband speech ($\SI{16}{\kilo\hertz}$).
It relies on linear prediction to reduce computational complexity, hence generating the signal in the residual linear prediction domain.
The decoding step is divided into two parts: a frame-rate network that computes the conditioning for every $\SI{10}{\milli\second}$ frame using the coded parameters, and a sample-rate network that computes the conditional sampling probabilities.
LPCNet predicts the new excitation sample using the previously generated excitation and speech samples, as well as the current linear prediction sample from the $16$th-order linear prediction.

More recent work~\cite{kleijn2021generative} presented a new neural speech decoder (Lyra) compressing speech at \kbps{3}.
The encoder directly codes stacked mel-spectra and the decoder uses noise suppression and variance regularization to improve the quality of out-of-distribution samples.
When compared to the proposed solution, Lyra is conditioned on a substantially different bit stream and works under different conditions (e.g. noisy speech).

To the best of our knowledge, there exists no prior GAN vocoder which allows frame-by-frame generation of speech at low delay or which provides high quality speech synthesis conditioned on a coded bit stream.

\section{Streamwise StyleMelGAN Vocoder (SSMGAN)}
\label{sec:architecture}
\subsection{Baseline StyleMelGAN}
\label{subsec:baseline_smgan}
StyleMelGAN~\cite{smgan} is a lightweight neural vocoder allowing synthesis of high-fidelity speech with low computational complexity. 
It employs Temporal Adaptive DE-normalization (TADE) to style a noise vector with the acoustic features of the target speech (e.g., mel-spectrogram) via instance normalization and elementwise modulation.
More precisely it learns adaptively the modulation parameters $\gamma$ and $\beta$ from the acoustic features, and then applies the transformation $\gamma\odot c+\beta$,
where $c$ is the normalized content of the input activation.
For efficient training, multiple random-window discriminators adversarially evaluate the speech signal analyzed by a set of Pseudo-Quadrature Mirror Filters (PQMF)~\cite{pqmf} filter banks, with the generator regularized by a multi-resolution STFT loss. 
All convolutions in StyleMelGAN are non-causal and run as a moving-average on sliding windows of the input tensors. 
This results in significant amount of algorithmic delay due to the deep hierarchical structure of the model. 
In the following, we describe major modifications to this baseline model that enable the generation at very low delay with different acoustic features for conditioning. 

\subsection{Streamwise Convolution}
There are two requirements to operate a convolutional model in streaming manner with low algorithmic delay. 
First, the dependency on future inputs to predict the current output should be as low as possible. 
We achieve this by enforcing all convolutions in StyleMelGAN to be causal so that the model has zero delay.
The second requirement is to generate the output frame by frame, as the new input information is available. 
This condition is fulfilled in StyleMelGAN by adding an internal memory buffer to the causal convolutions in inference mode, as described in~\cite{Rybakov2020} and illustrated in Figure~\ref{fig:conv_causal}.
Each causal convolution stores a buffer containing the last input samples used for generating the previous output frame, and then reused once the new input sample is available. 
By applying the above modifications to StyleMelGAN, we obtain \textit{Streamwise StyleMelGAN (SSMGAN)}, which is able to generate speech signals frame by frame with no delay between the input conditioning features and the output waveform.     

\begin{figure}[t]
  \centering
  \centerline{\includegraphics[width=0.95\columnwidth]{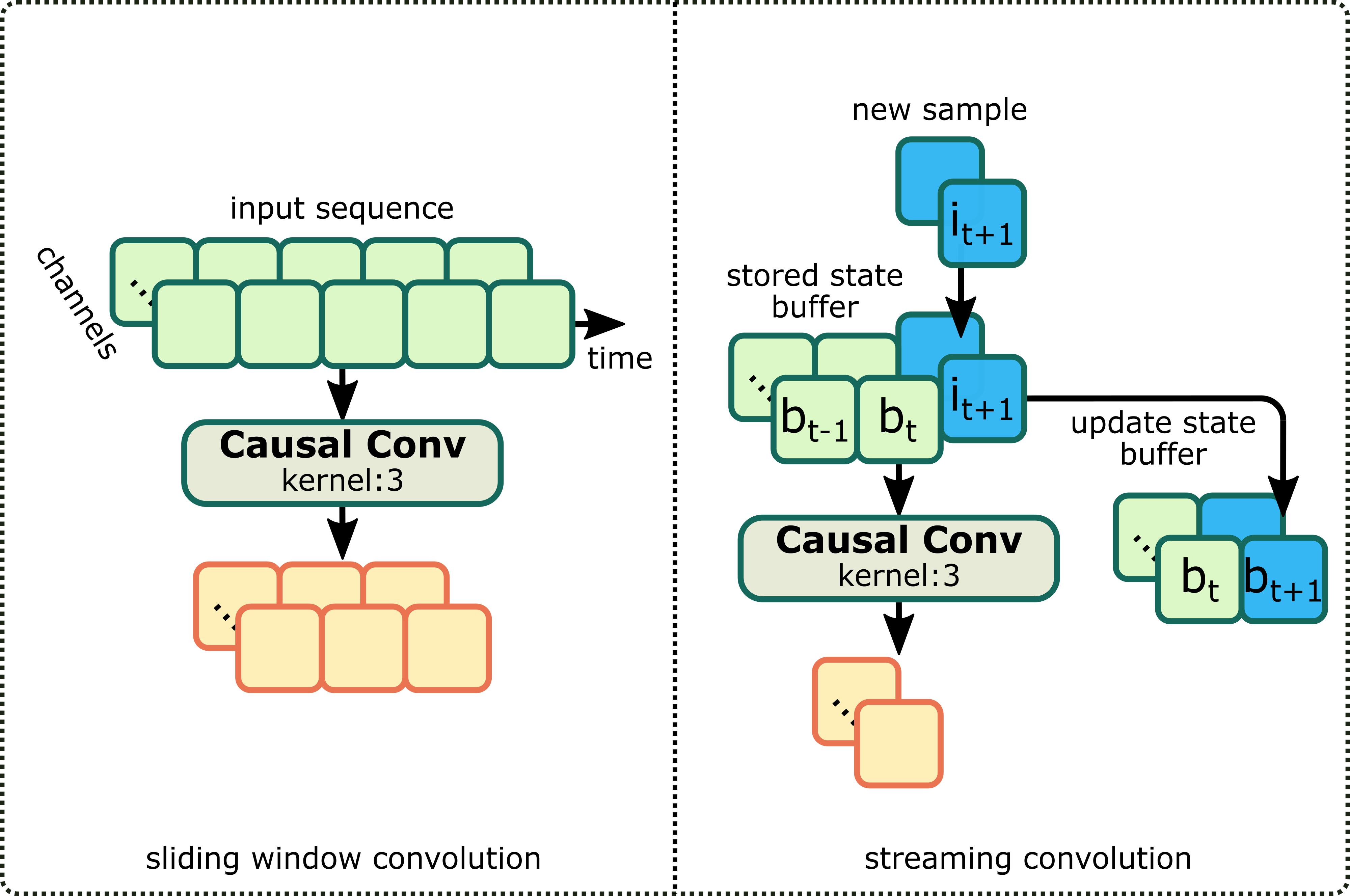}}
  \caption{Diagrams for non-streaming convolution (left) and streaming convolution (right)}
  \label{fig:conv_causal}
\end{figure}


\subsection{Channel Normalization}
It is not feasible to run instance normalization~\cite{ulyanov2016instance} in SSMGAN as the normalization statistics are estimated along the temporal dimension of the input activations.
We replace instance normalization with channel normalization~\cite{mentzer2020high}, that estimates the statistics along the channel dimension instead. 
Interestingly, we found this normalization maintains the model performance and keeps the training fast. 
It also avoids the creation of subtle clicking artifacts that sometimes occur when training StyleMelGAN with instance normalization on a multi-speaker dataset.  

\subsection{Modified TADE Residual Block}
The TADE residual blocks are slightly modified from the original model, as shown in Figure~\ref{fig:mod_taderes}.
The complexity in SSMGAN is reduced by using a single TADE conditioning layer and applying the same modulation parameters $\beta$ and $\gamma$ twice rather than having two separate TADEs in the residual block.
With this modification, the total number of model parameters reduces from \SI{3.86}{M} to \SI{2.73}{M}.

\begin{figure}[htb]
  \centering
  \centerline{\includegraphics[width=1.0\columnwidth]{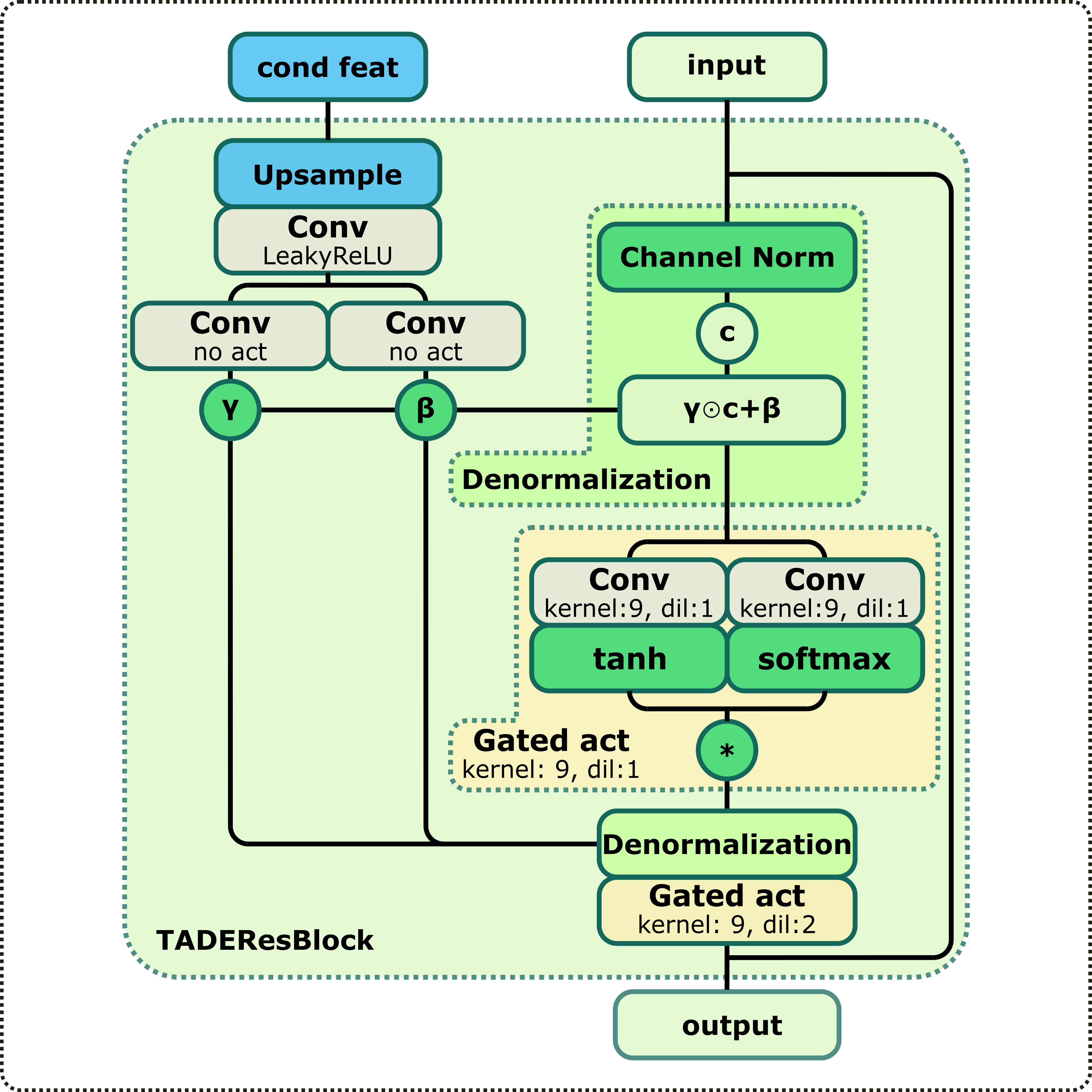}}
  \caption{Modified TADE residual block for the SSMGAN.}
  \label{fig:mod_taderes}
\end{figure}

\subsection{Multiband Generation}
SSMGAN further reduces the complexity compared to the baseline model by introducing multiband synthesis as in~\cite{Yu2020,Yang2020mb}.
Rather than synthesizing the whole band of the speech signal in time domain at the output sampling rate $f_s$, the generator outputs simultaneously different sub-bands sampled at $f_s/N$ Hz, with $N=4$ and $f_s=\SI{16}{kHz}$. 
By design, SSMGAN generates the sub-bands as an $N$-channels output, which is then fed to a PQMF synthesis filter-bank to obtain a frame of synthesized speech. 
Since the PQMF uses a filter prototype with 50\% of overlap, it incurs a delay of $1$ frame.

\subsection{Conditioning on Coded LPCNet Features}
\label{subsec:pitch_prior}
Finally, we condition SSMGAN with coded parameters in real-time to run as a speech decoder.
Instead of providing the mel-spectrogram as an intermediate representation, the coded parameters obtained by the LPCNet encoder at \kbps{1.6} are introduced to the generator network.
The pitch lag was found to be critical for high-quality synthesis, and hence it is processed separately from the rest of the conditioning information.
More precisely, the coded cepstral and energy parameters are passed through a simple causal convolutional layer to obtain an $80$ channel representation used for conditioning the generation from the prior signal.
This prior is not created from latent random noise, but rather from a learned embedding of the pitch lag which is then multiplied elementwise by the pitch correlation. 
Figure~\ref{fig:streamwise_smGan_generator} shows the complete architecture of the proposed SSMGAN conditioned on the LPCNet coded parameters. 
With this setting, SSMGAN can generate wideband speech frames of $\SI{10}{\milli\sec}$ length and total delay of $\SI{55}{\milli\sec}$, where $\SI{45}{\milli\sec}$ is introduced by the original extraction of the LPCNet coding packets, while $\SI{10}{\milli\sec}$ are added by the PQMF synthesis filter-bank.  

\begin{figure}[htb]
  \begin{minipage}[b]{1.0\linewidth}
      \centering
      \centerline{\includegraphics[width=1.0\linewidth]{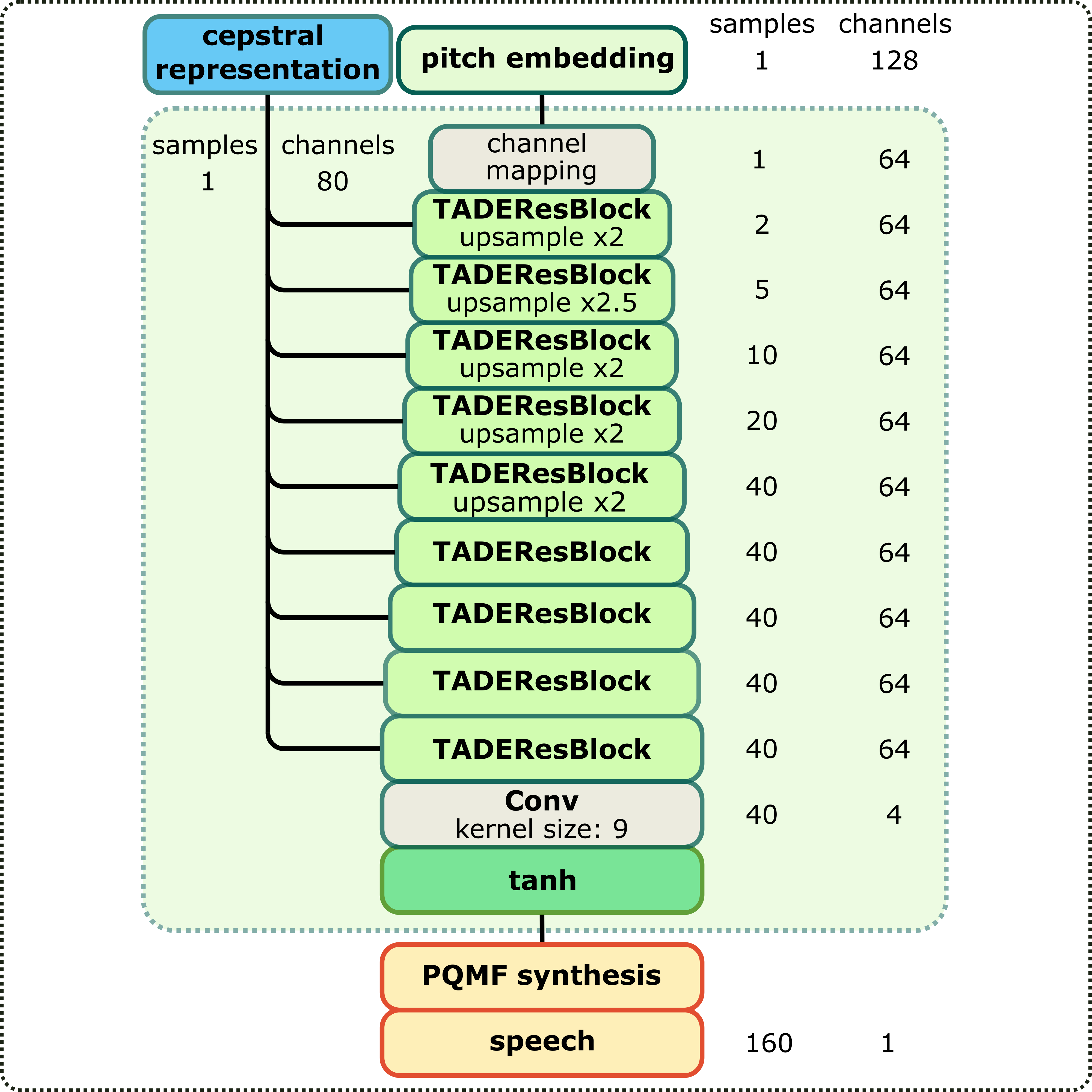}}
  \end{minipage}
  \caption{The SSMGAN generator.
  Dimensions are given for generating $1$ frame at \SI{16}{kHz} sampling rate. 
  The cepstral coefficients pass through a simple convolutional layer to obtain a representation of 80 channels.}
  \label{fig:streamwise_smGan_generator}
\end{figure}

\section{Experiments}
\label{sec:exps}
\subsection{Experimental setup}
The training procedure and hyperparameters are very similar to the ones described in~\cite{smgan}.
We train SSMGAN using one NVIDIA Tesla V100 GPU on the VCTK corpus~\cite{vctk} at \SI{16}{\kilo\hertz}.
The conditioning features are calculated as in~\cite{lpcnet} as described in Section~\ref{sec:rel_works}.
The generator is pretrained for \SI{200}{k} steps using Adam optimizer~\cite{kingma2014adam} with learning rate $lr_g=10^{-4}$, $\beta = \{0.5,0.9\}$.
When starting the adversarial training, we set $lr_g=5*10^{-5}$ and use the multi-scale discriminator described in~\cite{melgan} trained via Adam optimizer with $lr_d=2*10^{-4}$, and same $\beta$. 
The batch size is $32$ and for each sample in the batch we extract a segments of length $\SI{1}{\second}$.
The adversarial training lasts for about \SI{1.5}{M} steps.

\subsection{Subjective evaluation}
\label{subsec:evaluation}
We conducted a subjective listening test following the ITU-R MUSHRA~\cite{MUSHRA} recommendation comparing classical and neural speech coders.
The test set is composed of 12 utterances by 10 different speakers in 4 different languages.
All speakers and 3 out of 4 languages are unseen during training. Most of the utterances (10 out of the 12) are coming from unseen proprietary databases.
The obtained results with 16 expert listeners are shown in Figure~\ref{fig:mushra_smGan_1_6}.

\begin{figure}[htb]
  \begin{minipage}[b]{0.97\linewidth}
    \centering
    \centerline{\includegraphics[width=1.0\linewidth]{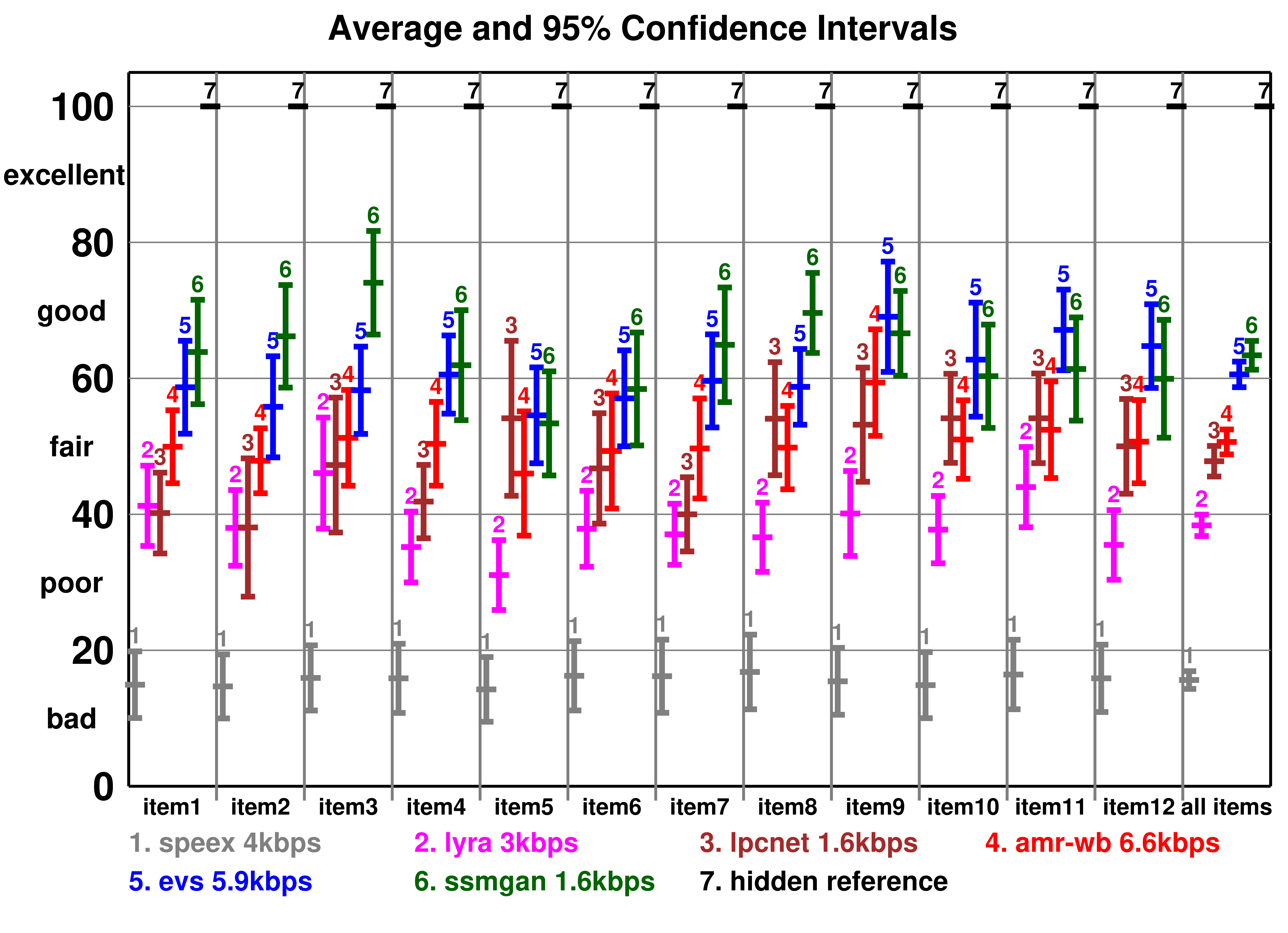}}
  \end{minipage}
  \caption{MUSHRA listening test results using $t$-distribution.}
  \label{fig:mushra_smGan_1_6}
  \end{figure}

The anchor is generated using the Speex speech decoder employed at a bit rate of \kbps{4}.
Two state-of-the-art neural decoders were considered: LPCNet at \kbps{1.6} and Lyra at \kbps{3}, as well as two classical but still widely used codecs: AMR-WB~\cite{amrwb} at \kbps{6.6} and the recent 3GPP EVS~\cite{evs} at \kbps{5.9}. The condition Lyra at \kbps{3} was generated using the release v0.0.1~\cite{lyra_code} with the default setting. 
EVS at \kbps{5.9} works with a variable bit rate (VBR) and that \kbps{5.9} reflects the average bit rate on active frames.
During a long inactive phase, EVS switches to a non-transmission mode (DTX), transmitting only periodically packets at a bit rate as low as \kbps{2.4}.
Since the test items only contain short pauses between sentences, the DTX mode plays a minor role in this test.

LPCNet was trained on the VCTK dataset.
One difference from the original work is that we do not apply a domain adaptation by first training on unquantized and then fine-tuning on quantized features, since this was found to make no difference on VCTK.
In addition, since VCTK is noisier and much more diverse than the NTT database used in the original work, we removed the data augmentation since it was found to be detrimental to the final quality\footnote{Check our demo samples at the following url: \url{https://fhgspco.github.io/ssmgan_spco/}}.
The publicly availabe version of the Lyra model was not retrained on VCTK, and hence it is not directly comparable with SSMGAN or LPCNet in this case.
It was nonetheless taken into consideration as it offers a reproducible benchmark.

\subsection{Objective evaluation}
\label{subsec:objective}
Our solution was also compared to the other neural decoders using different objective metrics.
Since it is known that objective speech quality models like POLQA~\cite{beerends2013perceptual} are not reliable for non-waveform-preserving coding schemes, and in particular for neural decoders, we also considered the newly introduced objective metric WARP-Q~\cite{jassim2021warpq}, which was designed for this purpose.
STOI~\cite{Taal11algorithmfor}, assessing the speech intelligibility, is also added, and the scores measured on 824 test items of VCTK are reported in Table~\ref{tab:objective}.

\begin{table}[h] 
  \centering
\begin{tabular}{lccc}
  \toprule
      Speech decoders & POLQA & STOI & WARP-Q \\
  \midrule
      Speex \kbps{4} & 2.022 & 0.720  & 1.074 \\
      AMR-WB \kbps{6.6} & 3.202  & 0.863 & \textbf{0.784} \\
      EVS \kbps{5.9} & \textbf{3.675} & \textbf{0.890} & 0.805\\
  \midrule
       LPCNet \kbps{1.6} & 2.628 & 0.777  & 0.915 \\
       Lyra \kbps{3} & 2.649  & 0.794 & 0.958 \\
       SSMGAN \kbps{1.6} & \textbf{2.719} & \textbf{0.830} & \textbf{0.826} \\
  \bottomrule
\end{tabular}
\caption{Average objective scores for neural decoders. For POLQA-MOS and STOI higher scores are better, while for WARP-Q lower scores are better (confidence intervals are negligible).}
\label{tab:objective}
\end{table}
SSMGAN at \kbps{1,6} scores the best among the neural coding solutions across all three metrics, which is in agreement with the subjective listening test.
The results of our MUSHRA listening test show moreover that these objective metrics do not fully reflect the perceived quality of the generated speech, disproportionately disfavouring generative models.

\subsection{Complexity}
\label{subsec:ccomplexity}
The main contribution to SSMGAN's computational complexity stems from the convolutions in the TADEResBlocks and the upsampling layers. 
If $L$ denotes the channel dimension, $K$ the size of the convolutional kernels, and $F$ the dimension of the input features, then (ignoring activations and lower order terms) the evaluation of a TADEResBlock takes $(F + 5L) L K$ multiply accumulate operations (MAC) per output sample.
Furthermore, an upsampling layer with kernel size $K$ and channel dimension $L$ takes $L^2 K$ MAC. 
With $L=64$, $K=9$, $F=80$ and TADEResBlock output sampling rates of $100, 200, 500, 1000, 2000, 4000, 4000, 4000,$ and $\SI{4000}{\hertz}$ this accumulates to
\begin{multline*}
(80 + 5\cdot 64)\cdot 64\cdot 9\cdot (100 + 200 + 500 + 1000 + 2000 + 4\cdot 4000) \\ + 64^2\cdot 9\cdot (200 + 500 + 1000 + 2000 + 4000) \approx \SI{4.8}{GMACs}.
\end{multline*}
A comparison with other neural vocoders used for neural speech coding is given in Table~\ref{tab:complexity}.
It should be noted, that the convolutional structure of SSMGAN allows for efficient parallel execution, which gives it a decisive advantage over autoregressive models on GPUs. 
The current unoptimized PyTorch implementation achieves about real-time frame-by-frame inference using four cores of an Intel(R) Core(TM) i7-6700 3.40GHz CPU.
The above complexity calculations show that the next step will be to work on an efficient implementation for mobile devices, which will be the object of a future work.

\begin{table}[h] 
  \centering
\begin{tabular}{ll}
  \toprule
       Model & Complexity \\
  \midrule
       SSMGAN (ours) & \flops{4.8} \\
       LPCNet~\cite{valin2019lpcnetspco} & \flops{1.5}\\
       Multi-band WaveRNN~\cite{Yu2020} & \flops{2.75} \\
  \bottomrule
\end{tabular}
\caption{Complexity of common neural vocoders for speech coding.}
\label{tab:complexity}
\end{table}

\section{Conclusion}
\label{sec:cond}
In this paper we introduce SSMGAN, a neural speech decoder generating state-of-the-art quality with low delay, complexity, and working at very low bit rate.
We assess the quality against existing neural autoregressive models and modern speech codecs at low bit rate, with both objective scores and subjective listening tests.
We show for the first time that GAN-vocoders can perform fast streaming speech synthesis with low algorithmic delay, and that they can achieve high quality synthesis when conditioned on compact parametric speech representations.

\bibliographystyle{IEEEtran}
\bibliography{waspaa_paper}

\end{sloppy}
\end{document}